%% file: gaussian.tex
\documentclass[a4paper,11pt]{article}
\usepackage{pos}
\usepackage{subcaption}
\usepackage[normalem]{ulem}

\title{Bottomonium spectral widths at nonzero temperature using maximum likelihood}
\author*[a]{Thomas Spriggs}
\author[a,b]{Gert Aarts}
\author[a]{Chris Allton}
\author[a]{Timothy Burns}
\author[c]{Benjamin J\"ager}
\author[d]{Seyong Kim}
\author[e]{Maria Paola Lombardo}
\author[a]{Sam Offler}
\author[a]{Ben Page}
\author[f]{Sin\'ead M. Ryan}
\author[g]{Jon-Ivar Skullerud}

\affiliation[a]{Department of Physics, Swansea University, Swansea SA2 8PP, United Kingdom}

\affiliation[b]{European Centre for Theoretical Studies in Nuclear Physics and Related Areas (ECT*) \& Fondazione Bruno Kessler Strada delle Tabarelle 286, 38123 Villazzano (TN), Italy}

\affiliation[c]{CP3-Origins \& Danish IAS, Department of Mathematics and Computer Science, University of
Southern Denmark, 5230 Odense M, Denmark}

\affiliation[d]{Department of Physics, Sejong University, Seoul 143-747, Korea}

\affiliation[e]{INFN, Sezione di Firenze, 50019 Sesto Fiorentino (FI), Italy}

\affiliation[f]{School of Mathematics \& Hamilton Mathematics Institute, Trinity College, Dublin 2, Ireland}

\affiliation[g]{Dept. of Theoretical Physics, National University of Ireland Maynooth, County Kildare, Ireland}

\emailAdd{t.spriggs.996870@swansea.ac.uk}

\abstract{
We present progress results from the {\sc Fastsum} collaboration's programme to determine the spectrum of the bottomonium system as a function of temperature using a variety of approaches.
In these proceedings, the Maximum Likelihood approach is used with an Ansatz comprising of a Gaussian spectral function for the ground state.
{\sc Fastsum} anisotropic lattices with $2+1$ dynamical quark flavours were used with temperatures ranging from 47 to 375 MeV.
}

\FullConference{%
 The 38th International Symposium on Lattice Field Theory, LATTICE2021
  26th-30th July, 2021
  Zoom/Gather@Massachusetts Institute of Technology
}

\begin{document}
\maketitle

\section{Introduction}

QCD changes dramatically as the temperature increases, from a confining hadronic phase to the quark-gluon plasma (QGP). Relativistic heavy-ion collision experiments probe this phase, but the interpretation of their results is clouded by the fact that there can be no detectors within the QGP itself, and hence properties of the high-temperature phase need to be indirectly derived from the hadrons observed in the low-temperature phase.

Charmonium and bottomonium mesons are particularly important in this regard because they can act as ``probes'' of the QGP. For example, the relative abundances of their excited states can be used as
a proxy for the QGP's temperature, as was proposed a long time ago \cite{Matsui:1986dk}.
The {\sc Fastsum} Collaboration has studied bottomonium using lattice QCD at nonzero temperature on anisotropic lattices for some time \cite{Aarts:2010ek,Aarts:2011sm,Aarts:2014cda}.
More recently, a variety of different methods are applied to extract the spectrum at nonzero temperature.  
In this contribution a Maximum Likelihood approach is used; other approaches include the Backus-Gilbert method \cite{Ben-Lat21} and Kernel Ridge Regression, a Machine Learning approach \cite{Sam-Lat21}.

\section{Spectral reconstruction method}

In-medium properties of bound states in QCD can be fully described by the spectral function  $\rho(\omega)$, associated with a channel $H$. This spectral function is a function of energy $\omega$, but we also expect it to behave non-trivially with temperature. The two most important properties of bound states -- the mass $M$ and width $\Gamma$ -- are easily extracted from $\rho(\omega)$ once it has been determined.

Lattice simulations calculate correlation functions $G_H(\tau; T)$ of operators in the channel $H$.
These are related to the corresponding spectral functions via
\begin{equation}
G_H( \tau;T) = \int_0^\infty \frac{d\omega}{2\pi}\, K(\tau, \omega; T) \rho_H( \omega; T).
\label{eq:kernel-equation}
\end{equation}
In the case considered here, the bottom quarks propagate nonrelativistically and can be described by the nonrelativistic approximation of QCD (NRQCD). The kernel function $K$ is then independent of $T$ and given by  
\begin{equation}
K(\tau, \omega; T) = e^{-\omega \tau}.
\label{eq:Fastsumrel-kernel}
\end{equation}
Eq.~\eqref{eq:kernel-equation} illustrates the ill-posed nature of the problem: 
$G(\tau)$ is known at a finite number ${\cal O}(10-100)$ points, whereas to adequately represent the continuous function $\rho(\omega)$ requires perhaps ${\cal O}(1000)$ points. The solution is then not unique \cite{hadamard}.

The approach used here to overcome this ill-posed problem is to assume an Ansatz for the spectral function,
\begin{equation}
\rho_{\text{Ansatz}}(\omega)
= \mathcal{A}_{\text{ground}}\;
         e^{- \frac{(\omega - M_{\text{ground}})^2}{2\sigma^2}}
\;+\; \mathcal{A}_{\text{excited}}\; \delta(\omega-M_{\text{excited}}).
\label{eq:Ansatz}
\end{equation}
In this Ansatz, the ground state is represented by a Gaussian function with mass $M_{\text{ground}}$ and width $\sigma$.
A Lorentzian could be used as well \cite{Larsen:2019bwy}, but the Gaussian allows a simple closed form
for the correlator $G(\tau)$ and captures the thermal broadening of the spectral line.
Also, at the level of precision available to us, there is little numerical difference between the two.
The excited states are modelled as a single $\delta$-function which is a pragmatic choice -- more sophisticated models with more parameters would have reduced predictive power.

Using Eq.~\eqref{eq:Ansatz} in Eq.~\eqref{eq:kernel-equation}, and taking the lower limit in Eq.~\eqref{eq:kernel-equation} to $-\infty$, gives the Ansatz
for the correlation function
\begin{equation}
G(\tau) = A_{\text{ground}}\;
  e^{-(M_{\text{ground}} - \frac{\sigma^2\tau}{2})\tau }
   + A_{\text{excited}}\; e^{-M_{\text{excited}} \tau}.
\label{eq:fit}
\end{equation}
The amplitudes $\mathcal{A}$ and $A$ are trivially related.  
A Maximum Likelihood method is used to fit the lattice correlation functions
to Eq.~\eqref{eq:fit} with the main aim being the determination of the ground 
state mass and width for each temperature considered.

\section{Lattice parameters}

We use our {\sc Fastsum} Collaboration's \textit{Generation 2L} ensembles which
were generated at a number of temperatures as listed in Table \ref{table}, with ${\cal O}(1000)$ configurations for each temperature.
Our lattices are $32^3\times N_\tau$ in extent, and are anisotropic with the spatial and temporal lattice spacings
$a_s = 0.0351(2)$ fm and $a_\tau = 0.1227(8)$ fm respectively.
This allows greater sampling rates of the temporal correlation functions, increasing the predictive power.
The pion mass is $m_\pi=236(2)$ MeV \cite{Cheung:2016bym,HadronSpectrum:2008xlg,Edwards:2008ja}.
For more details of the Generation 2L ensembles, see  Ref.~\cite{Aarts:2020vyb}.
 
\begin{table}[h]
 \centering
    \begin{tabular}{c||c|c|c|c|c|c|c|c|c|c|c}
    $N_\tau$ & 16 & 20 & 24 & 28 & 32 & 36 & 40 & 48 & 56 & 64 & 128 \\
    \hline
    T [MeV] & 375 & 300 & 250 & 214 & 187 & 167 & 150 & 125 & 107 & 94 & 47
    \end{tabular}
    \caption{Temporal extents and corresponding temperatures for the {\sc Fastsum} Generation 2L ensembles~\cite{Aarts:2020vyb}.
    The $N_\tau= 128$ ensemble was kindly provided by the Hadron Spectrum Collaboration
    \cite{Cheung:2016bym,HadronSpectrum:2008xlg,Edwards:2008ja}.
        }
    \label{table}
\end{table}

\section{Results}


\begin{figure}[t]
    \centering
    \includegraphics[width=0.9\textwidth]
    {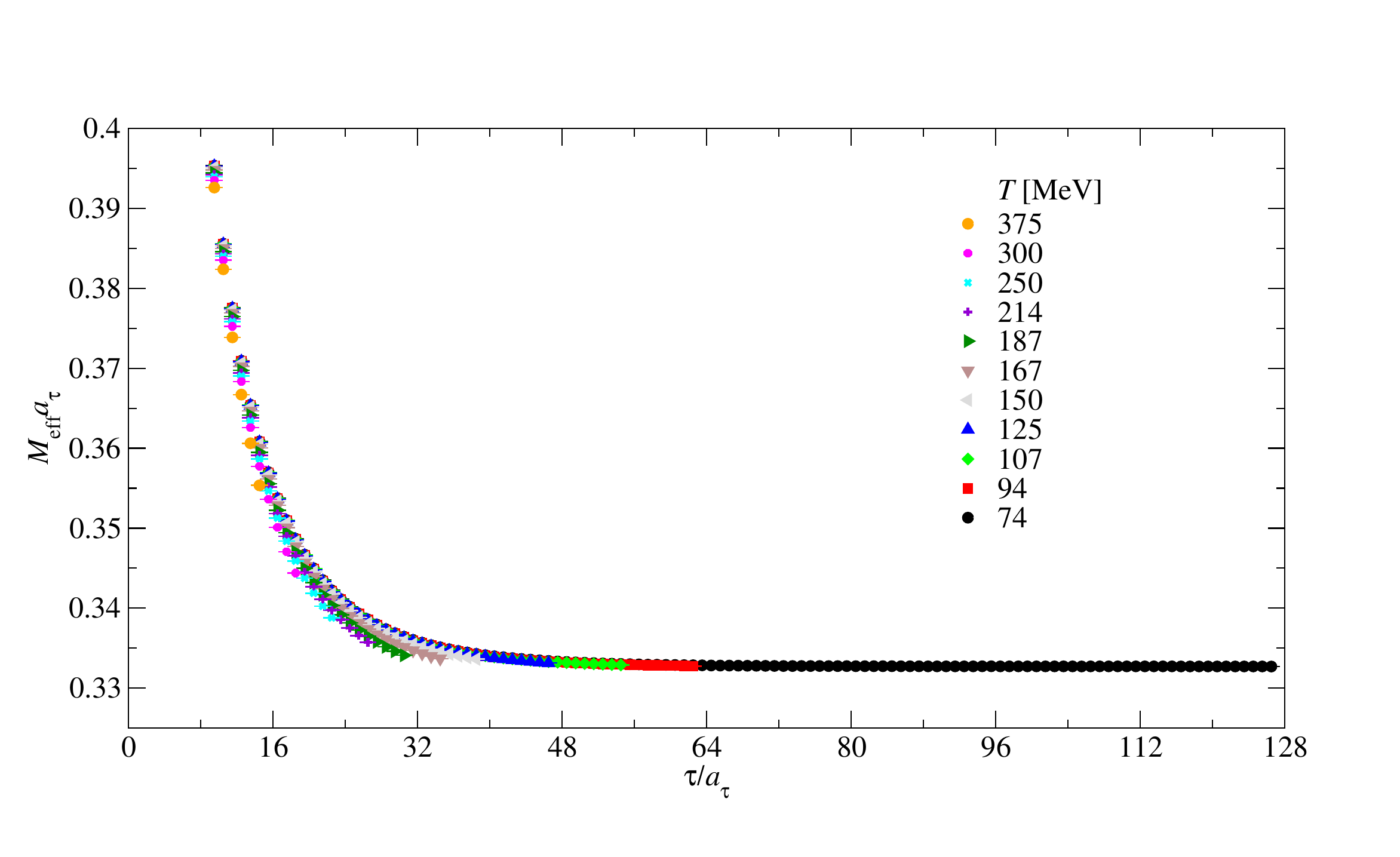}
    \caption{Effective mass plot in the $\Upsilon$ channel, for all temperatures considered. }
    \label{fig:example_eta_b_eff_mass}
\end{figure}

In Fig.~\ref{fig:example_eta_b_eff_mass} we plot the effective mass, defined by
\begin{equation}
M_{\text{eff}}(\tau) = \log \frac{G(\tau)}{G(\tau + 1)},
\label{eq:effective_mass}
\end{equation}
in the $\Upsilon$ channel, for all temperatures. 
As can be seen the contribution of the excited states is significant, so an accurate determination of ground state properties requires a careful choice of the time window $[\tau_1,\tau_2]$ included in the fit.
Ideally, to reduce the effect of the excited states, the time window would have $\tau_1, \tau_2 \rightarrow \infty$ with $\tau_2-\tau_1$ fixed.
In practice, we vary both $\tau_1$ and $\tau_2$ and study the dependencies of
the fit parameters $M$ and $\sigma$.
We find that varying $\tau_1$ requires some care because the chosen Ansatz requires there to be an excited state, and, from Fig.~\ref{fig:example_eta_b_eff_mass} this will not be the case for large $\tau_1$.
We will investigate this systematic further in Ref.~\cite{inpreparation}.

\begin{figure}[t]
\includegraphics[width=0.5\textwidth]{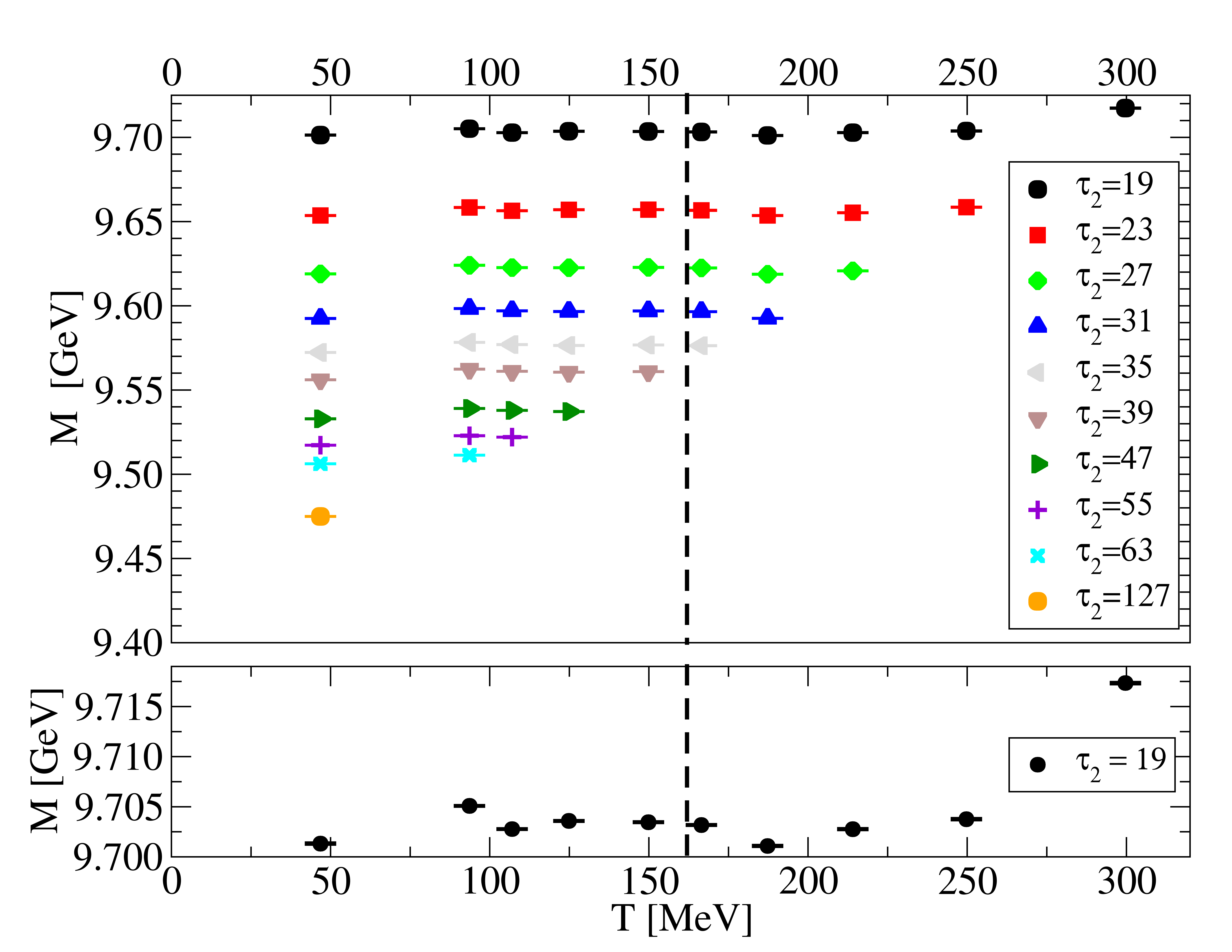}
\includegraphics[width=0.5\textwidth]{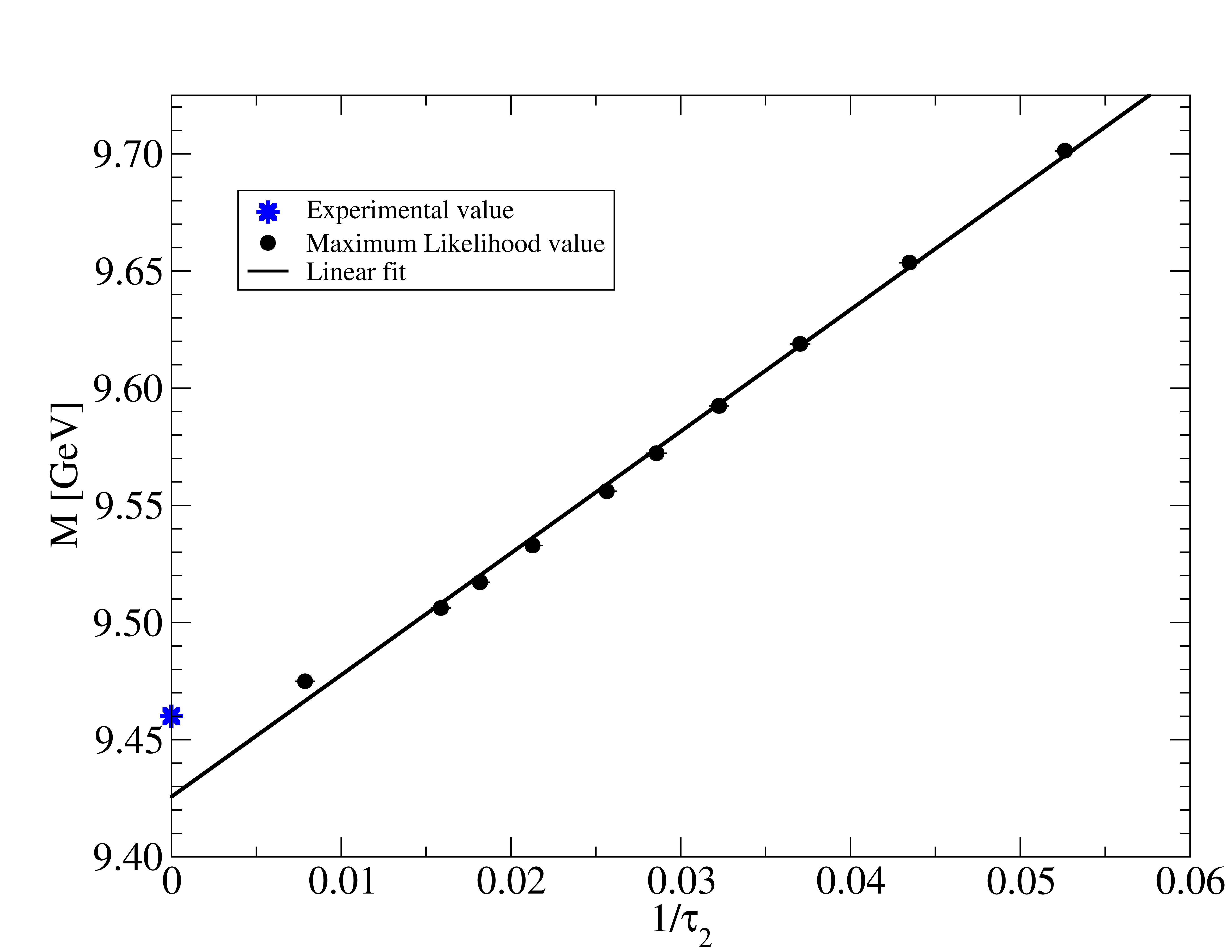}
\caption{
Left: Mass of the $\Upsilon$ state as a function of temperature for various $\tau_2$ values and $\tau_1=8$. The pseudo-critical temperature for our ensembles, $T_{\rm pc}=162(1)$ MeV \cite{Aarts:2020vyb}, is shown by vertical dashed line. The lower pane contains a close-up of the $\tau_2=19$ results for all temperatures.
Right: Mass for the $\Upsilon$ at $T=47$ MeV as a function of the inverse time window parameter $1/\tau_2$, with $\tau_1=8$ throughout. The black line shows an indicative linear fit to all points, and the star at $1/\tau_2=0$ indicates the experimental $\Upsilon$ mass.
}
\label{fig:test_against_t2}
\end{figure}

The systematic uncertainties associated with the $\tau_2$ choice are quite instructive.
In Fig.~\ref{fig:test_against_t2} we plot the mass obtained from our fit procedure for the $\Upsilon$ channel, as a function of temperature, for various $\tau_2$ and with
$\tau_1=8$ throughout.
In the top-left pane, the masses with the same $\tau_2$ values are plotted using
the same colour. In this way we can see that when there is no change in the
fitting procedure (i.e., no change in $\tau_1$ or $\tau_2$) the resulting
estimate of the mass shows little variation across a large range of temperatures.
However, there is some indication that for the largest temperature plotted,
i.e.\ $T=300$ MeV, the mass may increase compared to the lower temperatures; this will be investigated
in future work \cite{inpreparation}.
This is illustrated in the bottom-left pane which is a close-up of the
top-left pane showing only the $\tau_2=19$ fits, chosen because these fits are available for all $T\le 300$ MeV.

In the right pane of Fig.~\ref{fig:test_against_t2}, we plot the mass obtained for $T=47$ MeV as a function of $1/\tau_2$. This demonstrates the clear $\tau_2$ systematic effect.
The physical prediction is obtained in the $\tau_2 \rightarrow \infty$ limit, and a linear fit is shown to indicate this.

\begin{figure}[t]
\includegraphics[width=0.5\textwidth]{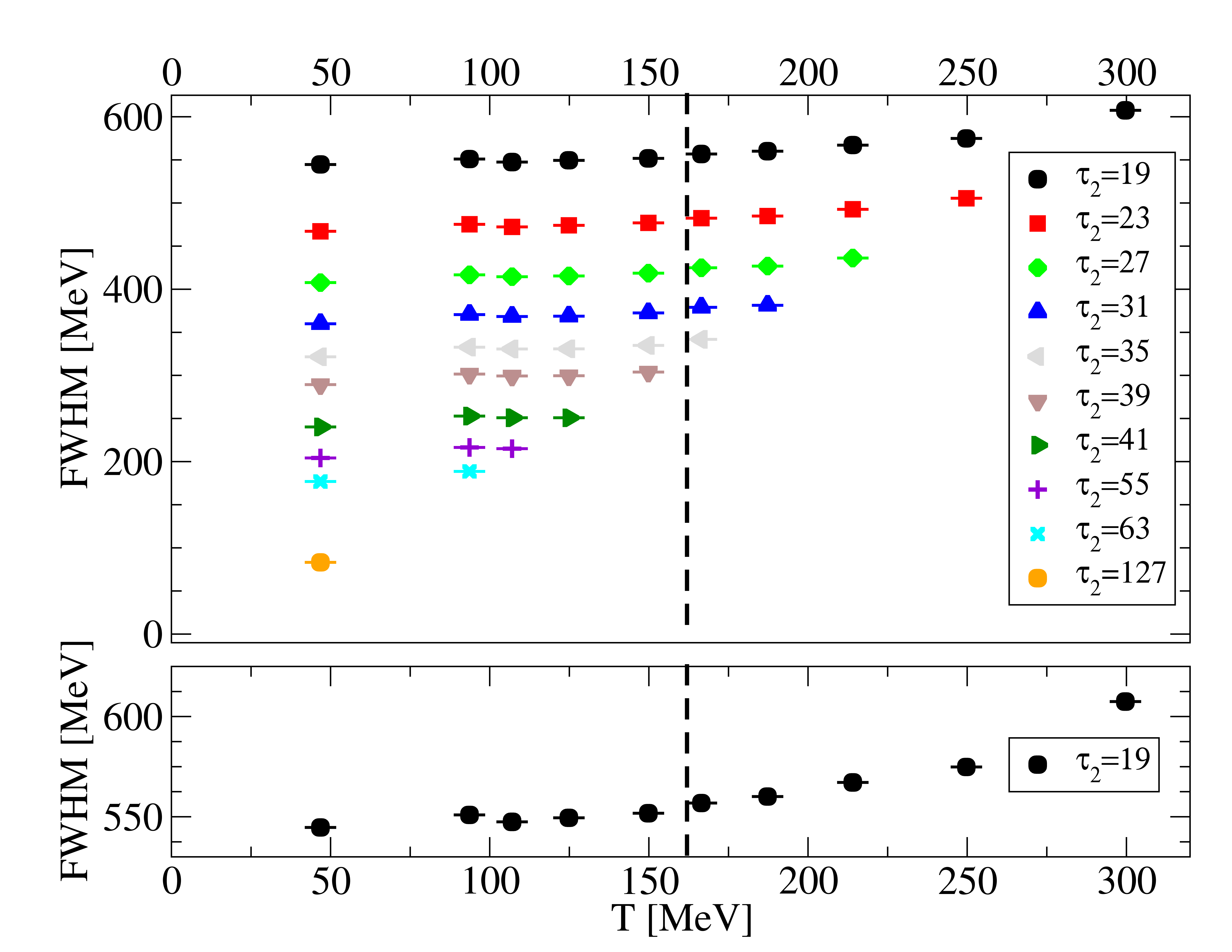}
\includegraphics[width=0.5\textwidth]{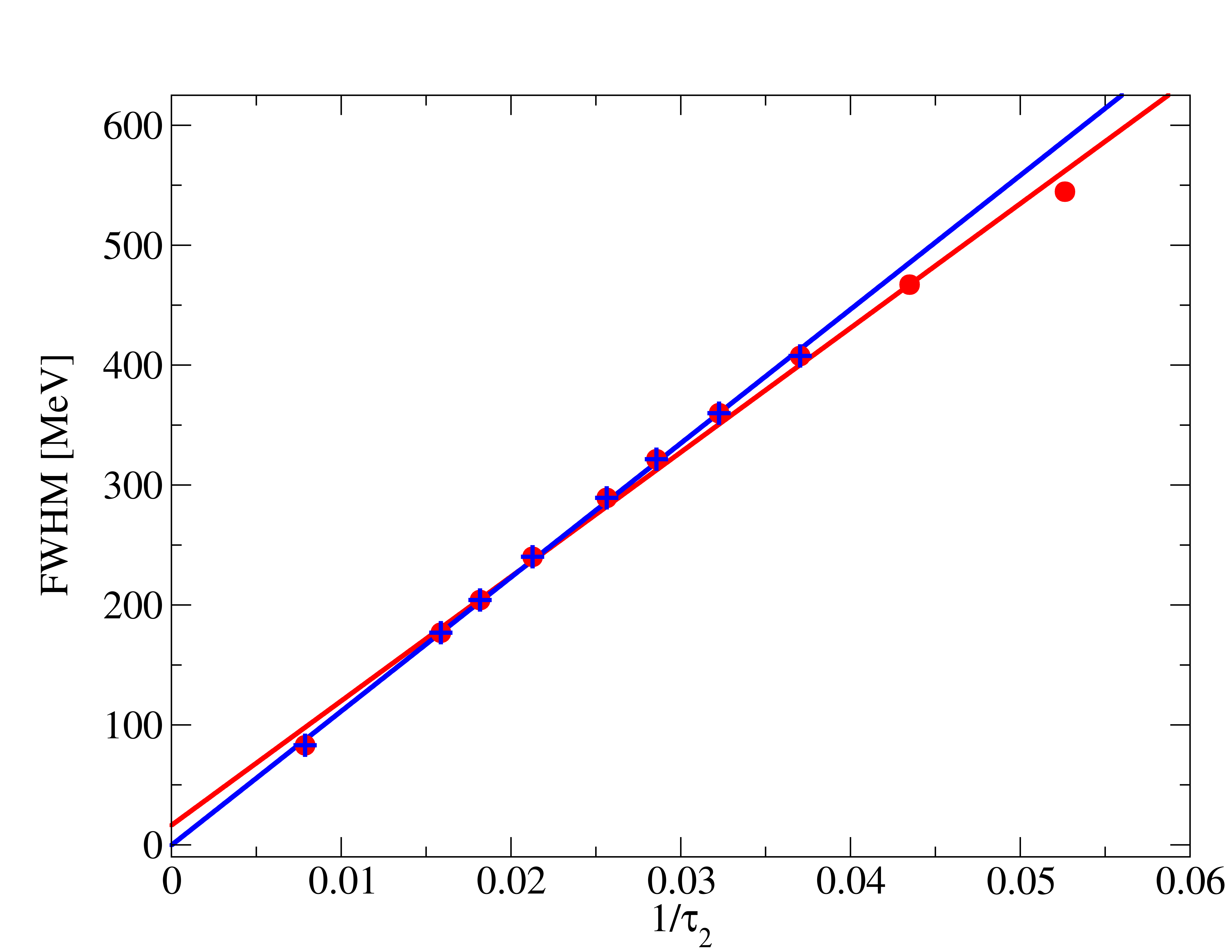}
\caption{
Left: FWHM of the $\Upsilon$ state as a function of temperatures for various $\tau_2$ values and $\tau_1=8$. The pseudo-critical temperature for our ensembles, $T_{\rm pc}=162(1)$ MeV \cite{Aarts:2020vyb}, is shown by vertical dashed line. The lower pane contains a close-up of the $\tau_2=19$ results for all temperatures.
Right: Full width at half maximum (FWHM) for the $\Upsilon$ at $T=47$ MeV as a function of the inverse time window parameter $1/\tau_2$, with $\tau_1=8$ throughout. The red (blue) line is a linear extrapolation on all (eight leftmost) data points.
}
\label{fig:tim_plot-mass}
\end{figure}

We now turn to the width. In Fig.~\ref{fig:tim_plot-mass} we plot the full width
at half maximum (FWHM), trivially obtained from $\sigma$.
The layout corresponds to Fig.~\ref{fig:test_against_t2}.
In the top-left pane the FWHM is plotted against $T$ for a various choices
of $\tau_2$.
Again the same $\tau_2$ values are plotted with the same colour. Clearly, in comparison to the mass, the width depends more strongly on $\tau_2$, with variations as large as a factor of five or so.
This illustrates how difficult it is to extract the width of a state
from a lattice correlation function using this approach.
Considering only the $\tau_2=19$ case, we see a variation in the width
above $T_{\rm pc}$.
This temperature dependent variation will be considered further in Ref.~\cite{inpreparation}.

In the right pane of Fig.~\ref{fig:tim_plot-mass} we plot the FWHM for $T=47$ MeV
against $1/\tau_2$.
Again this clearly shows the $\tau_2$ systematics in the width using our procedure.
We perform two linear extrapolations $1/\tau_2 \rightarrow 0$ indicated
by the red and blue lines.
The red line includes all points in this extrapolation,
and the blue line uses the eight left-most points only. 
Taking these extrapolations at face-value yields a width estimate of $\sim 10$ MeV.
Full results for other channels, including the $\eta_b$ and the P-wave states,
will be detailed in Ref.\ \cite{inpreparation}.

\section{Conclusion} \label{sec:conclusion}

We have presented preliminary results using the Maximum
Likelihood approach to determine the ground state
properties of the $\Upsilon$ state at finite temperature.
A Gaussian Ansatz for the ground state's spectral
function was used, together with a simple $\delta$-function
to approximate the excited states.
We find that the choice of the end point $\tau_2$ of the
time window has a larger effect on the predicted mass and
width than effects from changing temperature
from 47 to 375 MeV.
The systematic effect from the time window is especially
pronounced in the width, underlining the difficulty
of accurately determining this quantity from lattice
simulations.

Further work will be presented in Ref.~\cite{inpreparation},
where the $\eta_b$ and $P$-wave states will be studied.
This will also contain a comprehensive study of a variety
of spectral reconstruction methods.

\acknowledgments 
\input{acknowledgements.tex}

\bibliographystyle{JHEP}
\bibliography{ref.bib}

\end{document}

%% file: acknowledgements.tex
This work is supported by STFC grant ST/T000813/1.
SK is supported by the National Research Foundation of Korea under grant NRF-2021R1A2C1092701 funded by the Korean government (MEST).
BP has been supported by a Swansea University Research Excellence Scholarship (SURES).
This work used the DiRAC Extreme Scaling service at the University of Edinburgh, operated by the Edinburgh Parallel Computing Centre on behalf of the STFC DiRAC HPC Facility (www.dirac.ac.uk). This equipment was funded by BEIS capital funding via STFC capital grant ST/R00238X/1 and STFC DiRAC Operations grant ST/R001006/1. DiRAC is part of the National e-Infrastructure.
This work was performed using PRACE resources at Cineca via grants 2015133079 and 2018194714.
We acknowledge the support of the Supercomputing Wales project, which is part-funded by the European Regional Development Fund (ERDF) via Welsh Government,
and the University of Southern Denmark for use of computing facilities.
We are grateful to the Hadron Spectrum Collaboration for the use of their zero temperature ensemble.